\documentclass[12pt]{article}
\usepackage{epsfig}
\usepackage{amssymb}

\newcommand{\be}{\begin{equation}}
\newcommand{\ee}{\end{equation}}
\newcommand{\bea}{\begin{eqnarray}}
\newcommand{\eea}{\end{eqnarray}}

\newcommand{\kt}{\rangle}
\newcommand{\br}{\langle}

\newcommand{\ed}{\end{document}}

\newcommand{\pbr}{\prec\!}
\newcommand{\pkt}{\!\succ}

\newcommand{\bi}{\begin{itemize}}
\newcommand{\ei}{\end{itemize}}

\begin{document}

\title{Reply to Comment on ``Time-dependent quasi-Hermitian Hamiltonians
and the unitary\\ quantum evolution''}
\author{\\
Miloslav Znojil
\\
\\
\'{U}stav jadern\'e fyziky AV \v{C}R, 250 68 \v{R}e\v{z}, Czech
Republic\footnote{e-mail: znojil@ujf.cas.cz}}
\date{ }
\maketitle

\begin{abstract}

In his fresh ``Comment" (arXiv:0711.0137v1), A.~Mostafazadeh
reacts on my very recent letter (arXiv:0710.5653v1) where I tried
to clarify certain misunderstandings which occurred in A. M.,
Phys.~Lett.~B \textbf{650}, 208 (2007) [arXiv:0706.1872v2,
``Paper"]. As long as the ``Comment" offers a new support of the
original assertions made in the ``Paper", I feel obliged to
re-clarify the matter by extending my argumentation. I insist that
it is possible to escape the main conclusion of the ``Paper",
indeed. In particular, I point out a gap in the new calculations
in ``Comment", add a few remarks on the notation and reconfirm
that the unitarity of the time-evolution DOES NOT require the
time-independence of the metric operator.

\vspace{5mm}

\vspace{2mm}


\end{abstract}


 \noindent
In the notation used in my preprint \cite{znojil} as well as in
the A. Mostafazadeh's brand new comment on it \cite{comment}, the
symbol
    $\Theta$ denotes a  (positive) metric
    operator with a square root $\omega:=\sqrt\Theta$, both
    assigned to
     a possibly time-dependent $\Theta$-pseudo-Hermitian
    Hamiltonian operator $H$ acting on a reference Hilbert space
    ${\cal H}$ with the inner product $\br\cdot|\cdot\kt$.
Moreover,
 \begin{itemize}
    \item symbol $h:=\omega H\omega^{-1}$ is chosen to denote the
    equivalent Hermitian Hamiltonian leading to the evolution operator $u$
    such that
    $i\hbar\partial_t u(t)=h(t)u(t)$ and $u(0)=I$,
    where $I$ stands for the identity operator,
    \item
     the first of Eqs.~(17) of \cite{znojil}
     (rewritten as Eq. Nr. (3) in \cite{comment})
     {\em defines} the auxiliary
     quantity
        \be
        U_R(t)=\omega(t)^{-1}u(t)\omega(0),
        \label{e17}
        \ee
        \item
  in \cite{comment}, the subsequent unnumbered equation
  re-derives  Eq. Nr. (11) of ref.~\cite{plb},
    \be
    \Theta(t)=U_R(t)^{-1\dagger}\Theta(0)\:U_R^{-1}
    .\ee
    \end{itemize}
 \noindent
On this background the core of the new misunderstandings can be
easily spotted as lying in the {\em incorrect} assumption
represented by Eq. Nr. (2) in \cite{comment}, viz., by the
relation
        \be
        i\hbar\partial_t U_R(t)=H(t)U_R(t),~~~~U_R(0)=I\,.
        \label{U=}
        \ee
An easy explanation of the new puzzle is obtained when we
differentiate the definition~(\ref{e17}) and reveal that the
assumption (\ref{U=}) is manifestly incorrect. At the same time,
precisely this contradictory and entirely unfounded relation was
postulated in  \cite{plb} and used to derive the final statement
represented by the last Eq. Nr. (4) in ref.~\cite{comment}.

We can repeat that there emerges no obstacle which would violate
the unitarity of the quantum evolution when the Hamiltonian
$H=H(t)$ becomes manifestly time-dependent. In \cite{znojil} we
verified that in place of the puzzling Eq. Nr. (12) of
ref.~\cite{plb} (or in place of its unnumbered version preceding
eq. Nr. (4) in \cite{comment}) one has, simply,
        \be
        H(t)^\dagger=\Theta(t)H(t)\Theta(t)^{-1}.
        \label{ph}
        \ee
In the other words, one returns to the expected time-dependent
quasi-Hermiticity condition for $H(t)$.

This being said, it may sound paradoxical when we propose to
complement the above brief rebuttal to the Ali Mostafazadeh's {\em
technical} comment by an additional text expressing our {\em full
agreement} with his {\em philosophical} conclusion that ``The root
of the misjudgment made in [\ldots] seems to be the rather
deceptive nature of the notation". Let us add a few more remarks
on that matter, therefore.

We believe that probably the main source of the possible
ambiguities should be seen in the fact that our decision of
working with the quasi-Hermitian Hamiltonians $H$ (i.e., with
those which obey the operator identity (\ref{ph}) with a
nontrivial $\Theta \neq I$) {\em implies} that we have to work
with {\em several} Hilbert spaces {\em at once}.

In a brief detour let us note that in a historical perspective,
the work with several Hilbert spaces found its strong, persuasive
and {\em purely physical} original motivation in nuclear physics.
In a way reviewed by Scholtz et al \cite{Geyer} people often try
to {\em start} there from a {\em prohibitively complicated}
Hamiltonian operator $h$ which acts in a standard physical Hilbert
space ${\cal H}^{(stand)}_{phys}$ with the elements $\Phi$ marked,
for our present purposes, by the ``curly" ket symbols $|\Phi\pkt$.

All the textbook \cite{Messiah} wisdom applies: one can employ the
standard Dirac's notation and work, at least formally, with the
current spectral representation of the Hamiltonian,
 \be
 h = \sum_{n=0}^\infty\,|n\pkt\,E_n\,\pbr n|\,
 \label{spec}
 \ee
keeping in mind that for our $h=h^\dagger$ we are allowed to
assume, in the simplest scenario, that the basis $\{\,|n\pkt\,\}$
is orthogonal and complete in ${\cal H}^{(stand)}_{phys}$. Still,
in a more or less purely empirical manner, people found out that
there exist several different mappings of the original and exact
``complicated" $h$ on its various ``simpler" (though formally
equivalent) versions $H$.

In a way exemplified by the above-mentioned formula $h=\omega
H\omega^{-1}$, the second, different, ``reference" Hilbert space
${\cal H}={\cal H}^{(ref)}$ enters the scene. Typically, in
\cite{Geyer}, a ``realistic"  multinucleon Hamiltonian $h$ was
assigned a simpler, bosonic isospectral partner $H$. At present,
we may already read about a quickly growing number of applications
of this idea in several branches of physics (cf., e.g., the Carl
Bender's thorough review \cite{Carl} of the so called ${\cal
PT}-$symmetric models in field theory).

Leaving the applications and switching to an (in principle,
possible \cite{ali}) generalization of the mappings with $\omega
\to \Omega$ and $h=\Omega H\Omega^{-1}$, each invertible mapping
$\Omega$ seems to transform $h$ of eq.~(\ref{spec}) (considered as
acting on a given ket vector $|\Phi\!\pkt \ \in {\cal
H}^{(stand)}_{phys}$) into $H$. Thus, we can perceive the latter
operator as simply acting on a given ket-vector element of {\em
another}, reference space ${\cal H}= {\cal H}^{(ref)}$ as
mentioned above,
 \be
 |\Phi\kt := \Omega^{-1}\, |\Phi\pkt\ \in  {\cal H}^{(ref)}\,,
 \ \ \ \ \
 \br\Phi| :=\ \pbr \Phi|\,\left [\Omega^{-1}\right ]^\dagger
 \ \in  \ \left [{\cal
H}^{(ref)} \right ]^\dagger\ \sim\ {\cal H}^{(ref)}\,.
 \ee
It is worth noticing that by definition, {\em both } the old and
new spaces are {\em self-dual}. At the same time they are {\em
not} unitary equivalent since, by construction,
 \be
 \pbr \Phi|\Phi'\pkt\ = \br \Phi|\Theta
 |\Phi'\kt\,,\ \ \ \ \ \ \ \Theta=\Omega^\dagger\,\Omega=\Theta^\dagger>0\,.
 \label{zaved}
 \ee
We see that another Hilbert space has to be introduced. Here, it
will be denoted by the symbol ${\cal H}^{(\Theta)}$ (or simply by
${\cal H}_{phys}$ without superscript). It will {\em share} its
ket vectors with the ``intermediate" space ${\cal H}= {\cal
H}^{(ref)}$ (where the inner product was $\br \cdot|\cdot\kt$). In
parallel, it will {\em differ} from it by the innovated definition
of its linear functionals,
 \be
 {\cal H}_{phys}^\dagger
 :=\left \{
 \br\!\br \Phi| = \br \Phi| \Omega^\dagger\Omega \,\equiv\,
 \pbr\Phi| \,\Omega
 \ \
 \right \}\,.
 \ee
In this sense, only the mapping between ${\cal
H}^{(stand)}_{phys}$ and ${\cal H}_{phys}$ can be considered
norm-preserving and unitary.

We arrived at the very heart of the conflict between the differing
opinions concerning the notation conventions as expressed in
\cite{znojil} and \cite{comment}. We feel that the danger emerging
from a ``hidden subtlety" of the self-duality questions has simply
been underestimated in the latter text. Indeed, once you consult
any textbook \cite{Messiah} you immediately imagine that this
danger is in fact specific for all the situations where one works
with more (albeit unitarily equivalent!) representations of the
physical Hilbert space. Then, indeed, the (usually, trivial)
correspondence between ``the ket vectors" (i.e., the elements of
the space) and ``the bra vectors" (i.e., the linear functionals in
the same space) deserves an enhanced attention.

That's why we repeat our older recommendation \cite{znojil} that
the linear functionals in ${\cal H}_{phys} ={\cal H}^{(\Theta)}$
with $\Theta \neq I$ {\em should} be denoted by the {\em
specific}, doubled Dirac's brackets $\br\!\br \cdot|$.

\ed